# Electroplating of conformal electrodes for vacuum nanogap tunnel junction

L. Jangidze [a], A. Tavkhelidze [b, z], Y. Blagidze [c] and Z. Taliashvili [a]

[a] *Tbilisi State University, Chavchavadze ave. 13, 0179 Tbilisi, Georgia*

[b] *Ilia State University, Cholokashvili ave. 3/5, 0162 Tbilisi, Georgia*

[c] *Institute of Cybernetics, Sandro Euli st. 5, 0186 Tbilisi, Georgia*

[z] E-mail: avtotav@gmail.com

**Abstract**

In this study, we electroplate Cu electrode on Si substrate to realize a large-area vacuum nanogap for electron tunneling. We used cathode coating, cathode rotation, asymmetric current regime, and electrolyte temperature stabilization to obtain the regular geometry of the Cu electrode and reduce its internal tension. Subsequently, internal tension was altered to achieve the predefined surface curvature (concave or convex). For 12-mm diameter Ag/Cu electrode, we achieve the curvature of 40 nm/mm from the Ag side. Reduction of the electrode diameter to 3 mm allowed curvature as low as of 2.5 nm/mm. It also allowed fabrication of two conformal electrodes having a nanogap of less than 5 nm wide, over the area of 7 mm$^2$. Such electrodes can be used for efficient energy conversion and cooling in the mixed thermionic and thermotunneling regime.

**Introduction**

In recent years, micro-coolers and energy converters based on electron quantum-mechanical tunneling have been investigated [1–6]. The first study of cooling by electron tunneling has been done to avoid overheating in the single-electron transistors [1]. The typical metal-insulator-metal (MIM) tunnel junction has high heat conductivity owing to the low insulator thickness of 3–10 nm. This results in a parasitic heat backflow, which decreases the cooling coefficient. The metal-vacuum-metal (MVM) junction is free from this drawback, but at the same time is very complex from the standpoint of practical realization. However, a conformal electrode method based on the electrochemical growth of conformal electrodes has been proposed [2]. Coolers utilizing vacuum nanogap were studied in a number of earlier works [3–7]. It has been shown that MVM junction cooling power is of the order of 100 W/cm$^2$ [3, 5]. However, the cooling coefficient of such junctions does not exceed 10%. Further, a metal-vacuum-insulator-metal (MVIM) tunnel junction with an additional thin insulator coating layer has been studied [6]. It was found that the cooling coefficient of MVIM junction is of the order of 40–50% and, in addition, the insulating layer protects the electrodes against the short circuit and simplifies design. Composite electrode materials were also studied from the point of view of maximizing thermionic emission and thermotunneling [8]. Practical realization of energy converters and coolers based on thermotunneling has been discussed in the works [9-15].

The objective of this study is to find an electrochemical growth regime, which minimizes internal tension in the bulk Cu electrode grown on Si substrate. Internal tension must be low enough to obtain two conformal electrodes after separation of bulk Cu and Si. Matching of two surfaces should be accurate enough to allow electron tunneling in MVM junction formed between them.

**Experimental**

*Electrochemical growth.-* We used n-type Si(100) double-side polished substrates with diameter $D$=50 mm, 40 mm, and 20 mm, and thickness of 1–2 mm as a base electrode. Thin Ti and Ag (Ti – 0.1 µm and Ag – 1.2 µm) films were sputtered on the Si wafer in situ. Adhesion between the Ti and Ag films was artificially lowered. It was regulated precisely as described in [16]. Subsequently, the sample was exposed to the atmosphere and the thick Cu layer was electroplated on an Ag surface. Thick Cu layers were deposited in the sulfate electrolyte ($CuSO_4$ $5H_2O$ + $H_2SO_4$ + $C_2H_5OH$ + $H_2O$) in the thermo-stabilized bath, at the current density $J$=15–50 mA/cm$^2$, using mechanical stirring. The electrolyte was simple in its composition, stable and easy to correct and allowed high current densities. The ethyl alcohol was used to prevent the formation of one-valence Cu ions, thereby resulting in dense, finely crystalline precipitates. Electrically insulated current lead was connected to the Si wafer from the backside.

Electrolysis was carried out in the thermo-stabilized bath in the temperature $t$ range of $t$=23.5–35˚C. Temperature stabilization was used to exclude the influence of different thermal extraction coefficients of Si and Cu materials. Two baths were used to stabilize the electrolyte temperature. The internal bath filled with electrolyte was placed in the external bath filled with water. In the external bath, water temperature was maintained using heater and kept $\approx 2$˚C lower than in the internal bath. Electrolyte $t$ was varied in the range of 23.5–35˚C, and stabilized with an accuracy of 0.3˚C. Our first task was to obtain a uniform surface of the deposited Cu electrode. The following methods were used to control the Cu distribution over the cathode surface: (i) Coating of the cathode surface with the ring masks of electrically insulating material of different geometries. (ii) Placing additional cathodes near the main cathode. (iii) Cathode rotation. (iv) Use of different size anodes. (v) Using of asymmetric AC current.

Copper electrode with $D$=28 mm was deposited on 50-mm and 40-mm diameter Si wafers. In addition, $D$=12 mm and $D$=3 mm electrodes were deposited on 20-mm diameter wafers. The Cu thickness was 450–750 µm for 28mm and 12 mm electrodes, and up to 4000-µm for 3 mm diameter electrode. The Cu/Ag/Ti/Si sandwich (Fig. 1a) was

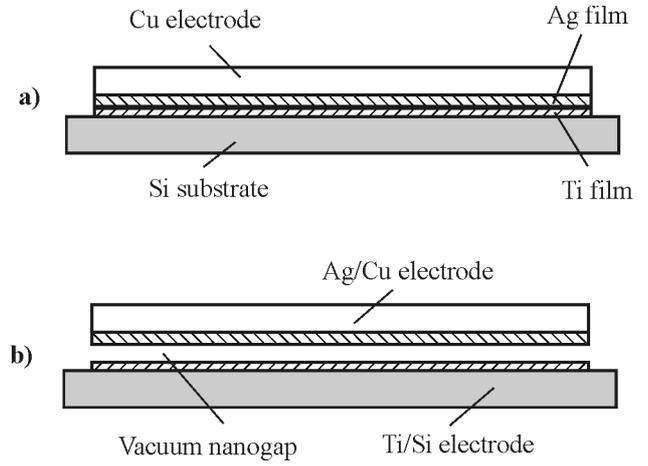

Fig. 1 a). Multilayer sandwich Cu/Ag/Ti/Si; b) Sandwich after cracking, Ti/Si and Ag/Cu electrodes are separated.

fabricated as described above. Subsequently, the sandwich was cracked using reduced adhesion between the Ti and Ag layers. Obtained bulk Ti/Si and Ag/Cu conformal electrodes are shown in Fig. 1b. Heating or cooling was used to crack the sandwich. Structure cracked owing to different thermal expansion of Si and Cu electrodes and low adhesion between the Ti and Ag layers. After cracking Ag/Cu electrode thickness was measured using a point micrometer (Heidenhain). Best samples were cracked inside the evacuated volume. Piezoelectric cylinders were used to maintain and regulate vacuum nanogap.

*Surface conformity measurement.-* Surface conformity (i.e. absolute curvature value) of Ag/Cu electrode was measured using a Michelson interferometer from the Ag side [17]. The He–Ne laser with the wavelength of $\lambda$=632.8 nm served as a light source. One of the interferometer mirrors were replaced by the mirror-surface of the Ag/Cu electrode. The interference pattern was formed in the air gap between one of the mirrors and a virtual image of the second mirror (sample in our case). When the gap was plain-parallel, fringes of equal inclination (circular fringes) were obtained (Fig. 2). After fixing, the sample was adjusted to obtain an interference pattern with equal inclination fringes. By counting the number of fringes $n$ the absolute curvature was calculated using the formula, $N = n \times \lambda/2$. The sign of the curvature was examined by pressing the longer leg of the mirror. Unbending of the fringes in the interference pattern indicated surface convexity and consequently, bending of the fringes indicated concavity. To determine a normalized curvature $\alpha$ of the metal electrode, the formula $\alpha = n(\lambda/2)/D$ was used.

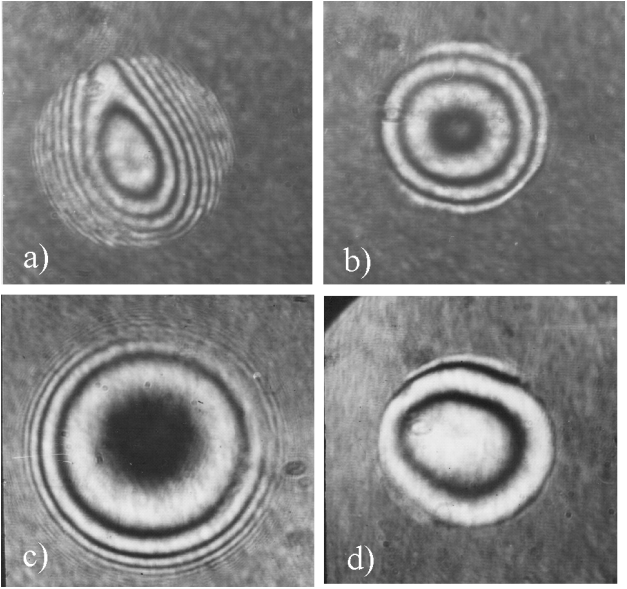

Fig. 2 a). $D$=12 mm Ag/Cu electrode (from the Ag side) deposited without cathode rotation; b) The same deposited with cathode rotation; c) $D$=20 mm Ti/Si elelectrode surface; d) $D$=12 mm Ag/Cu electrode surface (from the Ag side).

The initial normalized curvatures for base electrodes (Si wafers) of $D$=50 mm, 40 mm, and 20 mm were 57 nm/mm, 47 nm/mm, and 16 nm/mm, respectively. We monitored temperature during the interferometer measurements to keep it equal to the electrode growth temperature.

**Results and Discussions**

*Layer non-uniformity.*- In the early experiments, the Cu surface was non-uniform, with bulging edges (Fig. 3). The copper thickness at the edges was nearly twice as much as in the middle of the electrode. It has been demonstrated [18, 19] that the current density is distributed non-uniformly on the electroplated surface. The current density is higher in the corners and edges and lower in the middle area. Current passes along the field lines not normal to the surface as well as along the main field lines (normal to the electrode surface). To get rid of the edge defects we used protective masks of 3-mm thickness. Cu electrodes ($D$=28 mm) with thickness $h$ = 450–750 μm deposited the Ag/Cu electrode were concave with varying edge thicknesses (Fig. 4a).

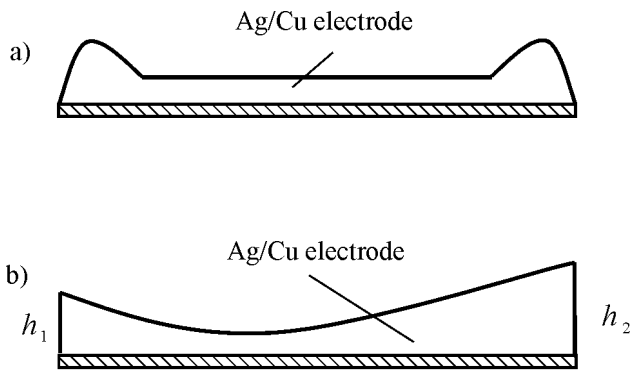

Fig. 3 Edges of 28 mm diameter Ag/Cu electrode grown without a protective mask (left) and with protective mask of 3-mm height (right).

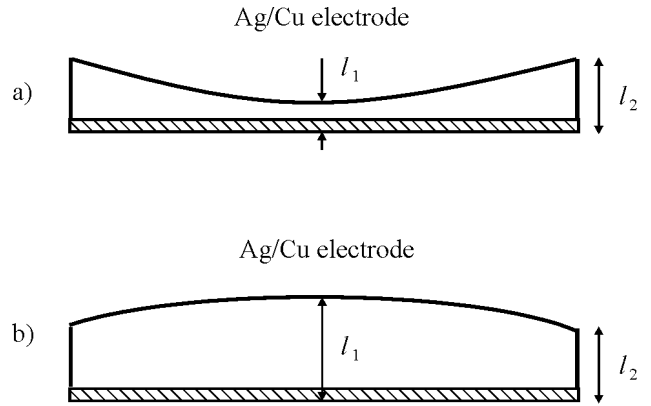

Fig. 4. a) Ag/Cu electrode grown with a protective mask of 3-mm height, b) Ag/Cu electrode deposited with cone mask and rotating cathode (walleye Cu surface); c) Ag/Cu electrode deposited with cone mask and rotating cathode (hill Cu surface).

Thickness in the lower part of the electrode $h_2$ has been always higher than that in the upper part of the electrode $h_1$. The average difference was $\Delta h_{av}$=105 μm. The Ag surface had a cylindrical shape with an average normalized curvature $\alpha_{av}$ =237 nm/mm.

To reduce the $\alpha$, 1 mm thick Si wafer diameter was reduced to 20 mm, and the Ag/Cu electrode diameter was reduced to 12 mm. To reduce the thickness difference $\Delta h_{av}$, the height of the ring insulator mask was increased to 10 mm. The Cu surface retained a similar concave shape with different edge heights, but the value of $\Delta h_{av}$ decreased to 41 μm. On the other hand, the Ag surface of the electrode had the same cylinder-like relief and $\alpha$ = 178 nm/mm (Fig. 2a). It is well known that the precipitates exhibit mechanical tension and it causes cathode curvatures [20]. To reduce the cathode $\alpha$, we decided to increase Si wafer thickness to up to 2 mm. As a result, $\alpha$ reduced and $\alpha$ =117 nm/mm was obtained from Ag/Cu electrode ($D$=12 mm). To reduce the non-uniformity of the edges of the electrode, additional cathode was placed near the main cathode. Additional ring electrode was placed on the outer diameter of the ring protective mask. The additional electrode attracted a fraction of the current, and substantially reduced the Cu thickness at the edges. The current density of the additional cathode was $J$=5 mA/cm$^2$. The Cu surface became effectively more uniform with $\Delta h_{av}$=9 µm ($N_{av}$=3.3 rings), and curvature reduced to $\alpha$=87 nm/mm (measured from the Ag side).

To get rid of surface non-uniformity, we rotated the cathode. Rotation also provided electrolyte steering. It resulted in uniform edges and central symmetrical curvature (Fig. 2b). In addition, the cone protective masks of 5 and 10 mm height, with hole diameters of 9/12, 10/12, 11/12, 12/12 mm, were used. Depending on the protective mask sizes and angular velocity of cathode rotation $V_\varphi$, concave (Fig. 4b) or convex (Fig. 4c) electrodes were obtained. In the case of 9/12 mm and 10/12 mm masks, the Cu surface was always convex, irrespective of the angular velocity. In the case of 11/12 mm and 12/12 mm masks at $V_\varphi$=2.5 rpm, the surface was essentially concave, whereas, at $V_\varphi$=10 rpm, it was convex. The average $\Delta h_{av}$, $\Delta l_{av}$, and $\alpha_{av}$ values are presented in Table 1. Here, $\Delta l$ is the thickness difference between the center and the edge (Fig. 4b). As Table 1 shows, the minimum $\alpha$ was obtained using the 10/12 mask.

Table 1. Values of $\Delta h_{av}$, $\Delta l_{av}$, and $\alpha_{av}$ in the case of $D$=12 mm electrodes with cone masks.

| Min/max hole D | 2.5 rpm | | 10 rpm | | $\alpha_{av}$ [nm/mm] |
|---|---|---|---|---|---|
| | $\Delta h_{av}$ [µm] | $\Delta l_{av}$ [µm] | $\Delta h_{av}$ [µm] | $\Delta l_{av}$ [µm] | |
| 9/12 | 10.5 | 115 | 12 | 149 | 126 |
| 10/12 | 3.7 | 45.5 | 5 | 89 | 84 |
| 11/12 | 7.4 | 9.4 | 10.7 | 75 | 210 |
| 12/12 | 4.1 | 46 | 5.9 | 39.5 | 195 |

Besides the standard-size Cu anodes (having an area twice as much as the cathodes), flat $D$=10 mm and tapered $D$=8 mm anodes were also used. In the case of $V_\varphi$=0 and with the standard protective mask of 10-mm height, we obtained $\Delta h_{av}$=13 µm. The introduction of taped masks and cathode rotation did not lead to substantial changes in this particular case.

To improve the surface curvature, we used the AC with different amplitudes of direct and reverse currents. In an earlier study [17, 21], 80–90% of the AC current period has been used to grow Cu, and the rest has been used to flatten the surface in the reverse current regime. We applied asymmetric AC with $J$=(0.35–1.15) mA/cm$^2$ and an asymmetry of 40–90% to the additional cathode at the frequency of 20, 30, and 50 Hz. For $V_\varphi$=0, we obtained a concave surface with $\Delta l_{av}$=36 µm and $\alpha$= 132 nm/mm.

To deposit the thick Cu layers (450–750 µm), we used asymmetric AC with $J$=(30–35) mA/cm$^2$ with the protective mask of 5-mm thickness and 12 mm hole diameter. In this case, the most uniform Cu surfaces with $\Delta l_{av}$=8.4 µm and the $\alpha$=158 nm/mm was obtained at $V_\varphi$=10 rpm.

*Internal tension.-* It is well known that rising electrolyte temperature accelerates deposition and introduces substantial changes. Precipitates become more plastic, which is important for curvature conservation. However, at higher $t$ cathode polarization decreases and macro-crystalline precipitates start to grow. Last can be compensated by increasing the current density and introducing electrolyte stirring [22]. To investigate the $\alpha(t)$ dependence of the 12-mm diameter Cu electrode it was grown using 10-mm height and 10/12 mm hole diameter cone protective mask. The cathode rotation speed was $V_\varphi$=10 rpm and the current density was $J$=(25–30) mA/cm$^2$. At $t$= 40–45°C, sandwiches split during the electrolysis process. To avoid splitting experiments were carried out within the limited range $t$= 23.5–35°C. The $\alpha(t)$ dependence for the 12 mm Ag/Cu electrode is illustrated in Fig. 5. The curvature decreases nearly

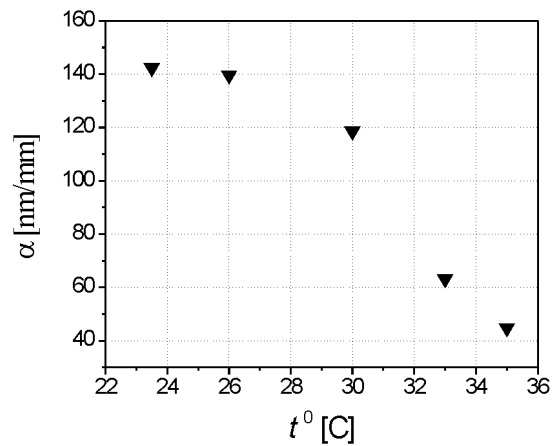

Fig. 5. $\alpha(t)$ dependence for $D$=12 mm Ag/Cu electrode.

three times with increasing *t* (interferograms from Ti/Si and Ag/Cu electrodes grown at *t*=35°C are also shown in Fig. 2). Next, we reduced the Cu electrode diameter down to 3 mm. Cu thickness was 700-2300 μm with $V_\varphi$= 2.5 rpm and $V_\varphi$=10 rpm. The current density was 15–20 mA/cm², and both the DC and AC regimes were used. Table 2 indicates that the best

Table 2. $\Delta h_{av}$, $\Delta l_{av}$, and $\alpha_{av}$ values for the Ag/Cu electrode with *D*=3 mm in the case of DC and AC.

|    | 2.5 rpm | | 10 rpm | | $\alpha_{av}$ [nm/mm] |
|----|---------|---------|---------|---------|---|
|    | $\Delta h_{av}$ [μm] | $\Delta l_{av}$ [μm] | $\Delta h_{av}$ [μm] | $\Delta l_{av}$ [μm] |    |
| DC | -       | -       | 3       | 6.7     | 74 |
| AC | 18.5    | 13.9    | 11      | 7.3     | 40 |

result $\alpha_{av}$=40 nm/mm, was obtained by applying the asymmetric AC with $V_\varphi$=10 rpm.

We also grew Cu of *D*=3 mm and up to 4000 μm thickness. In this case, the asymmetric AC density was increased to *J*=30–40 mA/cm² and Cu was deposited with $V_\varphi$=2.5 rpm, using 5-mm thick protective mask. Two extreme temperatures of 23.5°C and 35°C were investigated. The results are demonstrated in Table 3. The *t*-dependence was quite different for *D*=3 mm compared to *D*=12 mm. Curvature increased with increasing *t*. This can be ascribed to structural differences. We used high temperature and high current density to accelerate the growth of 4000-μm thick Cu. As growth speed increased, precipitates became of poor quality – fragile and porous, sometimes with dendrites. It is known from the literature [23] that when precipitate thickness increase grain size gradually increases and then saturates at some thickness. This leads to a reduction of the internal stress to constant value [18]. Considering this, we chose high current density *J*=30–40 mA/cm² at 23.5°C. This way, we reduced $\alpha$ down to 2.5 nm/mm, and in some cases, the samples without any detectable curvature were obtained.

Table 3. The $\alpha$ values of the 6 Ag/Cu electrode with *D*=3 mm for the electrolyte temperatures, 23.5°C and 35°C

| $\alpha$ [nm/mm] grown at *t*=23.5°C | <2.5 | <2.5 | 17.9 | 4.2 | <2.5 | <2.5 |
|---|---|---|---|---|---|---|
| $\alpha$ [nm/mm] grown at *t*=35°C | 105.4 | 7.3 | 4.2 | 8.4 | 34.8 | 3.2 |

*Electrode conformity-* The basic problem was the surface conformity of the two electrodes – Ti/Si and Ag/Cu. Initially, majority of Si substrates were convex (owing to mechanical polishing). Si wafer curve in one or another direction with respect to the initial curvature in the process of Cu growth. After cracking the wafer always restored its initial curvature. However, Ag/Cu electrode remains concave (Fig. 6a) or convex (Fig. 6b) depending on growth regime. Concave Ag/Cu electrode was needed to obtain the conformal pair. We investigate the dependence of the curvature sign (i.e. fraction of concave and convex surfaces in percentage terms) on the wafer thickness growth *t*. When wafers of 1-mm thickness was used, the percentage of concave surfaces were 22%. For 2 mm thick wafers the percentage of concave surfaces were 59%. Increase in wafer thickness enhances the number of concave surfaces to 2.7 times. Curvature sign dependence on *t* was investigated as well. The results for 2-mm wafer and electrode diameters of 12 mm and 3 mm are given in Table 4. The best results for the 12-mm and 3-mm Ag/Cu electrodes were obtained for the electrolyte temperatures 33°C and 35°C, respectively.

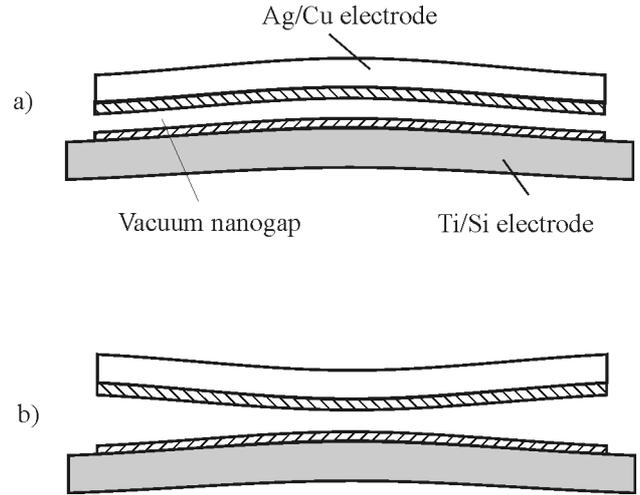

Fig. 6. Sandwich after cracking: a) Ti/Si electrode is convex and Ag/Cu electrode is concave; b) Ti/Si electrode is convex and Ag/Cu electrode is convex.

Table 4. A fraction of the concave surface of Ag/Cu electrodes having diameters of 12 and 3 mm at different *t*.

|   | Ag/Cu with *D*=12 mm | | | | | Ag/Cu with *D*=3 mm | |
|---|---|---|---|---|---|---|---|
| *t* (°C) | 23.5 | 26 | 30 | 33 | 35 | 23.5 | 35 |
| Concave [%] | 54 | 24 | 26 | 80 | 82 | 100 | 85.7 |

*Tunneling area evaluation.-* Further, we evaluated the tunneling area. It was defined as a region where the distance between the electrodes is $a<5$ nm. The tunneling area of the conformal electrodes (i.e. Ti/Si surface is convex and Ag/Cu surface is concave) can be estimated from simple geometric calculations. We consider the cross-section of the sandwich (Fig. 7). The tunneling area is the area of the ring with an external radius equal to the sample radius OC, and the internal radius OF. Replacing AC and BC arcs by straight lines we found that the contact area could be expressed by the formula: $S = \pi r^2 (a/H)(2 - a/H)$ with sufficient accuracy. Here, $H$=AB is the maximum difference between the curvatures and $a$=DE. The results for $D$=3 mm samples are given in Table 5. The $H$ was determined from $H = n \times \lambda$, where $n$ was the number of the measured fringes.

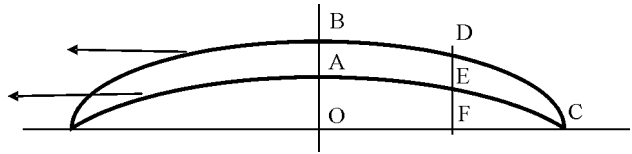

Fig. 7. Sandwich cross-section: AEC – Ti/Si electrode surface, BDC – Ag/Cu electrode surface.

Table 5. The number of interference rings and the corresponding contact area for the Ag/Cu electrodes with D=3 mm.

|  | Ag/Cu with $D$=12 mm | | | | | Ag/Cu with $D$=3 mm | |
|---|---|---|---|---|---|---|---|
| $t$ (°C) | 23.5 | 26 | 30 | 33 | 35 | 23.5 | 35 |
| Concave [%] | 54 | 24 | 26 | 80 | 82 | 100 | 85.7 |

Table 5 shows that the highest tunneling area were obtained for the $t$ = 23.5°C. For these samples, the whole area of the electrodes is contact area, signifying that the vacuum nanogap has a width of < 5 nm on the whole area of the $D$=3 mm electrodes.

## Conclusions

Cu electrodes were electroplated on Si substrate to achieve a large-area vacuum nanogap. To minimize the thickness difference between the electrode centre and edges we used 10 mm height protective masks. Cathode rotation allowed us to obtain central-symmetric electrodes and change the curvature sign in the desired direction. Increasing the thickness of the Si wafer from 1 mm to 2 mm allowed us to reduce the normalized curvature $\alpha$ by a factor of 1.5 and increase the output of the concave Ag/Cu electrodes by 37%. Increasing the electrolyte temperature to 35°C we reduced $\alpha$ down to 44.8 nm/mm, and improved output of the concave surfaces by 28%. Further, $\alpha$ was reduced down to 2.5 nm/mm by reducing the electrode diameter to 3mm and using the asymmetric AC. All surfaces were concave for growth temperature $t$=23.5°C. In this regime samples with no measurable electrode surface curvature (<2.5 nm/mm) were obtained. The tunneling area calculated on the basis of the measured parameters had maximum value of 7 mm$^2$.

To obtain conformal electrodes thick Si substrate should be used and the thick Cu layer should be grown with protective mask and cathode rotation. Cu electrode relative curvature decreases with increasing electrolyte temperature, increasing current density and reducing the Cu electrode diameter. Experiments demonstrate that electroplated surfaces can be used as electrodes in the vacuum nanogap devices.

## Acknowledgments

The authors thank M. Tetradze and T. Devidze for support in experiments. Work has been supported by Borealis Technical Ltd.